\documentclass[aps,twocolumn,groupedaddress,showpacs]{revtex4}
\usepackage{graphicx}
\begin{document}

\title{Dynamical structure factors of $S=1/2$ two-leg spin ladder systems}

\author{Nobuyasu Haga and Sei-ichiro Suga}
\affiliation{Department of Applied Physics, Osaka University, Suita, Osaka 565-0871, Japan}
\date{\today}
\begin{abstract}
We investigate dynamical properties of $S=1/2$ two-leg spin ladder systems. 
In a strong coupling region, an isolated mode appears in the lowest excited states, while in a weak coupling region, an isolated mode is reduced and the lowest excited states become a lower bound of the excitation continuum. 
We find in the system with equal intrachain and interchain couplings that due to a cyclic four-spin interaction, the distribution of the weights for the dynamical structure factor and characteristics of the lowest excited states are strongly influenced. The dynamical properties of two systems proposed for ${\rm SrCu_2O_3}$ are also discussed.

\end{abstract}
\maketitle
%
$S=1/2$ two-leg spin ladder systems with antiferromagnetic interactions have attracted great attention both theoretically and experimentally \cite{dagotto}. 
Fascinating aspects of elementary excitation as well as thermodynamic properties have been revealed. 
Using several theoretical methods, it was shown that $S=0$ and $S=1$ two-triplet bound-states exist below the two-triplet continuum in addition to the one-triplet excitation \cite{koto1,koto2,sach,jb,moni1,moni2,uhrig}. 
Such an $S=0$ two-triplet bound-state was identified in the optical conductivity measured for an $S=1/2$ two-leg spin ladder material ${\rm (Ca, La)_{14}Cu_{24}O_{41}}$ \cite{legex}. 
In this experiment, the strength of the interchain coupling ($J_{\bot}$) and the intrachain coupling ($J_{\|}$) was estimated as $J_{\|}/J_{\bot} \sim 1-1.2$ with $J_{\bot} \sim 1020-1100 { \rm cm}^{-1}$ \cite{legex,SCES}.  

In addition to the interchain and intrachain couplings, a cyclic four-spin interaction, which acts among four $S=1/2$ spins forming a plaquette, has been introduced to explain experimental findings for cuprate two-leg spin ladder systems. The analysis of the one-triplet mode observed by inelastic neutron-scattering experiments for ${\rm La_{6}Ca_{8}Cu_{24}O_{41}}$ revealed that a cyclic four-spin interaction is necessary to reproduce the observed dispersion relation \cite{ins,cyc}. 
For a $S=1/2$ two-leg spin ladder material ${\rm SrCu_2O_3}$, two sets of coupling constants, (i) $J_{\bot}=86 {\rm meV}, J_{\|}=172 {\rm meV}$ \cite{Johnson} and (ii) $J_{\bot}=150 {\rm meV}, J_{\|}=195 {\rm meV}, J_{cyc}=18 {\rm meV}, J_{diag}=3 {\rm meV}$ \cite{mizuno1} with $J_{cyc}$ and $J_{diag}$ being the coupling constants for a cyclic four-spin interaction and a diagonal interaction, have been proposed to reproduce temperature dependence of the susceptibility. 
It seems thus difficult to decide the proper model only from temperature dependence of the susceptibility. 
Detailed information on dynamical properties and low-lying excitations is desirable to discuss characteristics of $S=1/2$ two-leg spin ladder systems.

In this paper, we calculate the dynamical structure factor (DSF) of $S=1/2$ two-leg spin ladder systems using continued fraction method based on Lanczos algorithm.  Dynamical properties of $S=1/2$ pure two-leg spin ladder systems have already been studied in Refs. 15, 3, 16, 17, and 4. We perform systematic calculation for DSF and characteristics of the lowest excited states, laying stress on the effects of a cyclic four-spin interaction.

Let us first consider the $S=1/2$ two-leg spin ladder systems described by the following Hamiltonian, 
\begin{equation}
{\cal H} = J_{\|}\sum_{i=1}^{N/2} \left(
            \mbox{\boldmath$S$}_{1,i} \cdot \mbox{\boldmath$S$}_{1,i+1} 
          + \mbox{\boldmath$S$}_{2,i} \cdot \mbox{\boldmath$S$}_{2,i+1} 
                                  \right)
          + J_{\bot}\sum_{i=1}^{N/2} 
            \mbox{\boldmath$S$}_{1,i} \cdot \mbox{\boldmath$S$}_{2,i}, 
\label{eqn:s1}
\end{equation}
%
\noindent
where $\mbox{\boldmath$S$}_{l,i}$ denotes the $S=1/2$ spin operator in the $i$ th rung of the $l=1, 2$ chain, and $N$ is the total number of the site. 
The periodic boundary condition is applied along the chain. 
The energy is measured in units of $J_{\|}$. 
The DSF can be expressed in the form of a continued fraction \cite{GB87} as 
%
\begin{eqnarray}
S^{\mu}(q_x,q_y;\omega) 
   &=& \langle \Psi_{0}| S_{q_x,q_y}^{\mu \dagger} 
       \frac{1}{z-{\cal H}} S_{q_x,q_y}^{\mu} |\Psi_{0} \rangle 
                \nonumber \\
   &=&  S^{\mu}(q_x,q_y) C^{\mu}(q_x,q_y;\omega), 
   \hspace{4mm}(\mu = +, -, z) \nonumber \\
   &&\makebox[1cm]{}
\label{eqn:s2-2} 
\end{eqnarray}
%
\noindent
where $|\Psi_{0}\rangle$ is the eigenfunction corresponding to the lowest eigenvalue $E_0$, $q_x$ and $q_y$ denote the wave numbers along the leg and the rung, respectively, and $z=\omega+i\varepsilon+E_{0}$ with $\hbar =1$. We set $\varepsilon = 3.0 \times 10^{-2}$. 
The Fourier transform of the spin operator $S _{q_x,q_y}^\mu$ is given by 
$S_{q_x,q_y}^\mu = (1/\sqrt{N}) \sum_{l,j}
\exp[{\rm i}(q_x j + q_y l)] S_{l,j}^\mu. $
$S^{\mu}(q_x,q_y)$ is the static structure factor and $C^{\mu}(q_x,q_y;\omega)$ is represented in the form of the continued fraction. 
Since rotational symmetry around the $x, y$, and $z$ axes remains, we only calculate $S^z (q_x,q_y;\omega)$.

In Figs. 1(a) and 1(b), we show $S^{z}(q_x,\pi;\omega)$ and $S^{z}(q_x,0;\omega)$ for $N=28$, respectively, in $J_{\bot}/J_{\|}=0.5, 1$, and $2$. 
The weight of $S^z (q_x,q_y;\omega)$ is proportional to the area of the full circle. 
The accuracy of the continued fraction becomes worse with decreasing $J_{\bot}/J_{\|}$ and/or increasing $\omega$. 
In $J_{\bot}/J_{\|}=0.5$ the convergences have the relative errors about $O(10^{-3})$ for $\omega \geq 4$ and about $O(10^{-10})$ for $\omega < 3$, whereas in $J_{\bot}/J_{\|}=2$ they have the relative errors about $O(10^{-10})$ even for $\omega \sim 6$. 

\begin{figure}[htbp]
\begin{center}
\includegraphics[trim=0 0.6cm 0 0,clip,width=\linewidth]{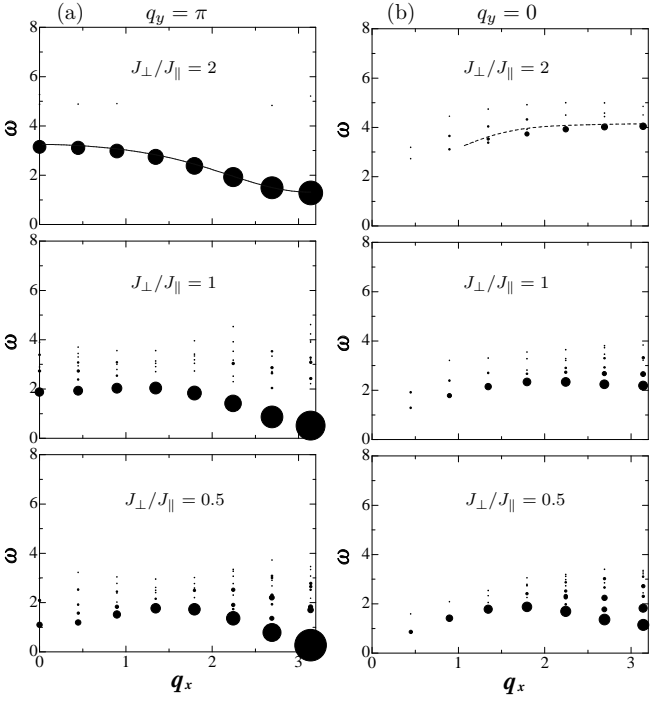}
\caption{
$S^z (q_x,q_y,\omega)$ for $S=1/2$ two-leg spin ladder systems in (a) $q_y=\pi$ and (b) $q_y=0$ for $N=28$. Parameters used are shown in the figures. 
The intensity of each pole is proportional to the area of the circle. 
The solid and broken lines represent the dispersion relations of the one-triplet excitation \cite{koto2} and $S=1$ two-triplet bound-state \cite{moni2}, respectively, adequate for $J_{\bot}/J_{\|} \gg 1$. 
}
\end{center}
\end{figure}

In Fig. 1(a), the largest weights in given $q_x$ lie in the lowest excited states, and the largest weight at $q_x=\pi$ becomes dominant among them with decreasing $J_{\bot}/J_{\|}$. The solid line represents the dispersion relation of the one-triplet excitation adequate for $J_{\bot}/J_{\|} \gg 1$ \cite{koto2}. 
The numerical results and the solid line agree with each other quantitatively. 
To discuss characteristics of the lowest excited states, we investigate the finite-size effects of the poles and their residues of the continued fraction $C^z (q_x,q_y;\omega)$~\cite{taka}. 
As reported in Refs. 19-21, a pole which belongs to a continuum tends to have appropriate size dependence at least either of its position or of its residue. 
In Fig. 2(a), we show the size dependence of the residue for the lowest excited states and that of the pole for the lowest and the second lowest excited states. In $J_{\bot}/J_{\|}=2$, no size dependence of $C^z (q_x,\pi;\omega)$ is seen in the lowest excited states, while clear size dependence appears in the pole for the second lowest excited states. The results show that the lowest excited states form an isolated one-triplet mode below the excitation continuum. 
In $J_{\bot}/J_{\|} = 1$, the size dependence is slightly seen around $q_x=0$ in the residue. Thus, the lowest excited states in $q_x \sim 0$ may be a lower bound of the excitation continuum, while those in the other wave numbers form an isolated mode. 
In $J_{\bot}/J_{\|}=0.5$, the residues show the size dependence in $0 \leq q_x \leq \pi$, indicating that the lowest excited states become a lower bound of the excitation continuum.
In $J_{\bot}/J_{\|}<1$, the elementary excitation may be described by two spinons in either chain dressed by a weak interchain interaction. 
The weights in $J_{\bot}/J_{\|} < 1$ are caused by such dressed spinons. 

\begin{figure*}[bt]
\begin{center}
\includegraphics[trim=0 0.8cm 0 0,clip,width=0.9\linewidth]{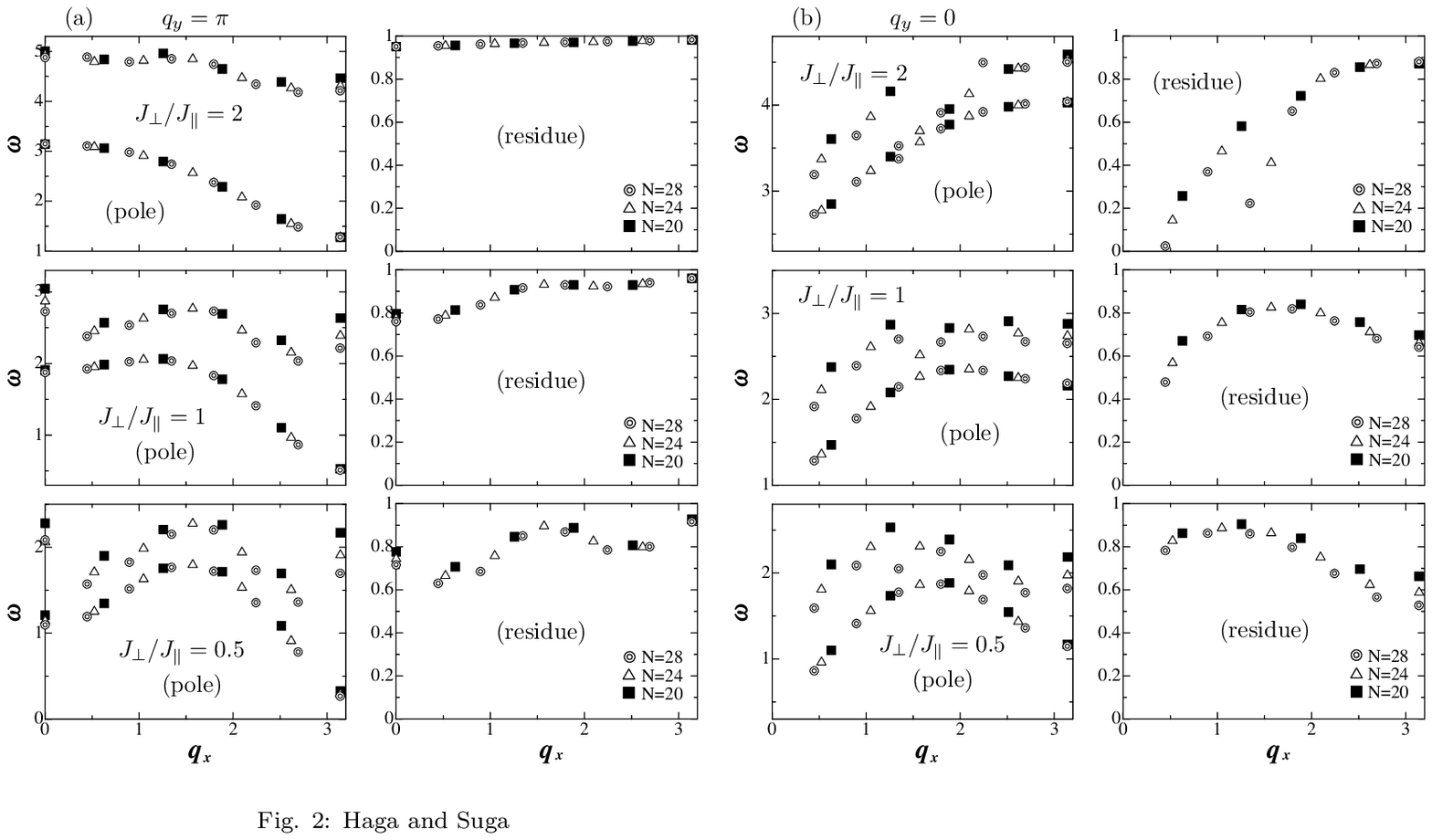}
\caption{
The size dependence of the poles and their residues corresponding to the results in Fig. 1. 
}
\end{center}
\end{figure*}
%

In Fig. 1(b), the broken line in $J_{\bot}/J_{\|}=2$, which represents the dispersion relation of the $S=1$ two-triplet bound state obtained for $J_{\bot}/J_{\|} \gg 1$ \cite{moni2}, agrees with the numerical results. 
Also in $q_y=0$, the largest weights in given $q_x$ lie in the lowest excited states.
As shown in Fig. 2(b), the residue in $J_{\bot}/J_{\|}=2$ decreases drastically around $q_x \sim 0.6\pi$ with decreasing $q_x$. This result means that the weights spread in many excited states in $q_x < 0.6\pi$. 
Therefore, in $J_{\bot}/J_{\|}=2$, the lowest excited states in $q_x > 0.6\pi$ form an isolated mode of the $S=1$ two-triplet bound state, while those in $q_x < 0.6\pi$ become a lower bound of the two-triplet continuum. In $J_{\bot}/J_{\|} \leq 1$, the size dependence of the residue becomes noticeable except for the region $q_x \sim 0$. Thus, the lowest excited states in $J_{\bot}/J_{\|} \leq 1$ are the lower bound of the continuum formed by dressed spinons. 

\begin{figure}[ht]
\begin{center}
\includegraphics[trim=0 0.7cm 0 0,clip,width=\linewidth]{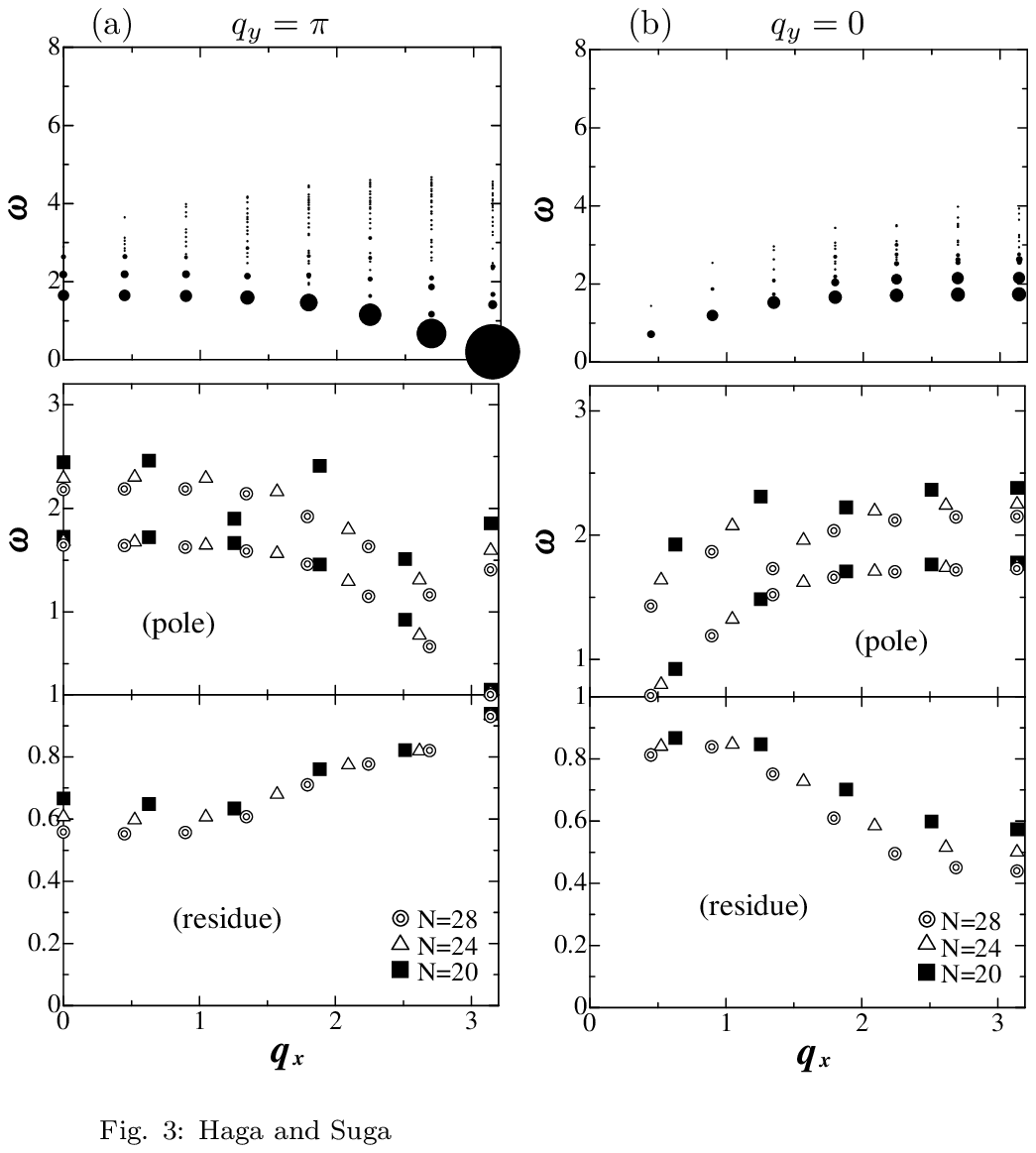}
\caption{
$S^z (q_x,q_y,\omega)$, and the size dependence of the poles and their residues  in $J_{\bot}/J_{\|}=1$ and $J_{cyc}/J_{\|}=0.1$ for $N=28$. 
}
\end{center}
\end{figure}

\begin{figure}[hbt]
\begin{center}
\includegraphics[trim=0 0.8cm 0 0,clip,width=\linewidth]{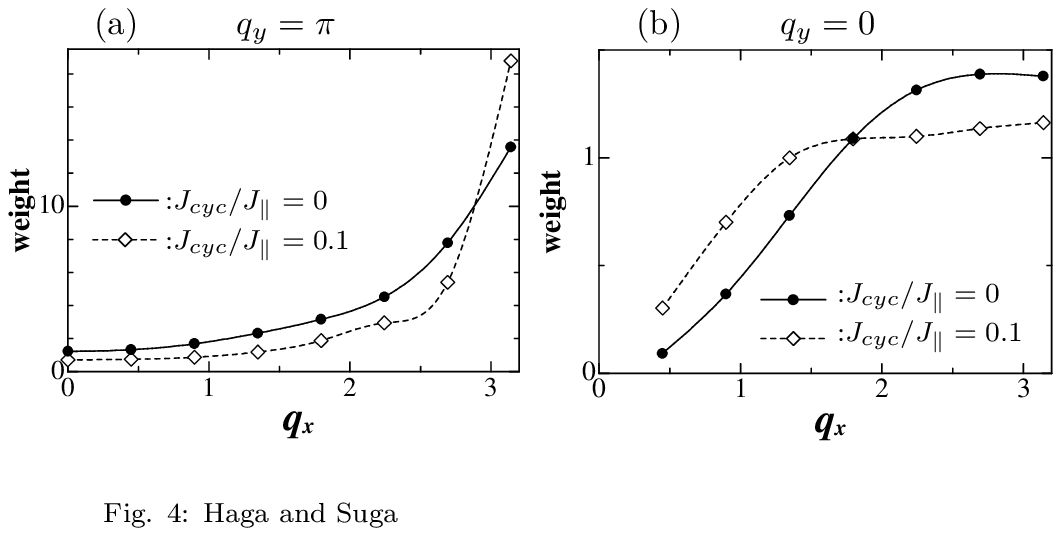}
\caption{
The weights of the lowest excited states for $J_{cyc}/J_{\|}=0.1$ and $0$ in $J_{\bot}/J_{\|}=1$. 
}
\end{center}
\end{figure}

\begin{figure}[htbp]
\begin{center}
\includegraphics[trim=0 0.7cm 0 0,clip,width=\linewidth]{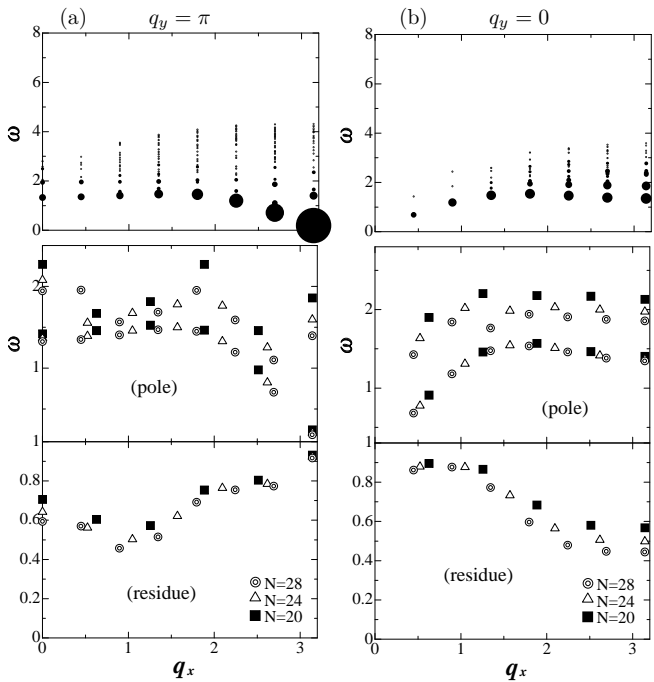}
\caption{
$S^z (q_x,q_y,\omega)$, and the size dependence of the poles and their residues  in $J_{\bot}/J_{\|}=0.770, J_{cyc}/J_{\|}=0.0923$, and $J_{diag}/J_{\|}=0.0154$ for $N=28$. 
Those values correspond to the parameter set (ii) for ${\rm SrCu_2O_3}$ \cite{mizuno1}. 
}
\end{center}
\end{figure}

%
\begin{figure}[hbt]
\begin{center}
\includegraphics[trim=0 1cm 0 0,clip,width=\linewidth]{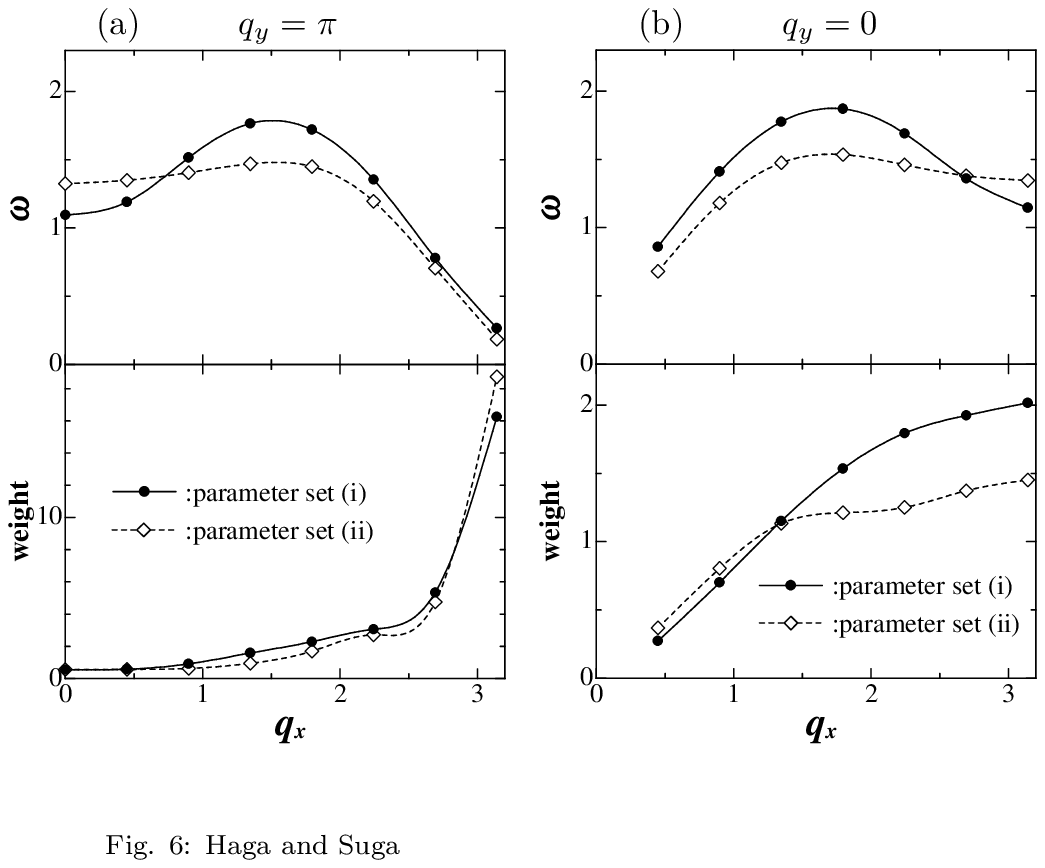}
\caption{
The dispersion relations and the weights of the lowest excited states in the parameter sets (i) and (ii) for ${\rm SrCu_2O_3}$. 
}
\end{center}
\end{figure}

We now investigated the effects of a cyclic four-spin interaction described by the following expression, 
\begin{eqnarray}
{\cal H}_{cyc} &=& J_{cyc}\sum_{i=1}^{N/2} \{ 4[
(\mbox{\boldmath$S$}_{1,i} \cdot \mbox{\boldmath$S$}_{1,i+1})
(\mbox{\boldmath$S$}_{2,i+1} \cdot \mbox{\boldmath$S$}_{2,i}) \nonumber \\
&&+ 
(\mbox{\boldmath$S$}_{1,i} \cdot \mbox{\boldmath$S$}_{2,i}) 
(\mbox{\boldmath$S$}_{1,i+1} \cdot \mbox{\boldmath$S$}_{2,i+1}) \nonumber \\
&&- 
(\mbox{\boldmath$S$}_{1,i} \cdot \mbox{\boldmath$S$}_{2,i+1})
(\mbox{\boldmath$S$}_{1,i+1} \cdot \mbox{\boldmath$S$}_{2,i})] \nonumber \\
&&+
(\mbox{\boldmath$S$}_{1,i} \cdot \mbox{\boldmath$S$}_{1,i+1}) + 
(\mbox{\boldmath$S$}_{1,i+1} \cdot \mbox{\boldmath$S$}_{2,i+1}) \nonumber \\
&&+
(\mbox{\boldmath$S$}_{2,i+1} \cdot \mbox{\boldmath$S$}_{2,i}) + 
(\mbox{\boldmath$S$}_{2,i} \cdot \mbox{\boldmath$S$}_{1,i}) \nonumber \\
&&+
(\mbox{\boldmath$S$}_{1,i} \cdot \mbox{\boldmath$S$}_{2,i+1}) + 
(\mbox{\boldmath$S$}_{1,i+1} \cdot \mbox{\boldmath$S$}_{2,i}) + \frac{1}{4}
\} . 
\end{eqnarray}
%
From inelastic neutron-scattering experiments for ${\rm La_{6}Ca_{8}Cu_{24}O_{41}}$, the coupling constants were estimated as $J_{\bot}=J_{\|}=110 {\rm meV}$ and $J_{cyc}=16.5 {\rm meV}$ \cite{ins}. Thus, we set $J_{\bot}/J_{\|}=1$ and $J_{cyc}/J_{\|}=0.1$. 
Combining the Hamiltonian (1) with the contribution (3), we calculate DSF. The results for $N=28$ are shown in Fig. 3. 
In the poles and their residues for $q_y=\pi$, noticeable size dependence comes out in contrast with the case of $J_{\bot}/J_{\|}=1$ and $J_{cyc}=0$ shown in Fig. 2. 
Therefore, in $q_y=\pi$ the isolated mode is reduced by a cyclic four-spin interaction, since it may act as a frustration term: The lowest excited states in $q_x<0.7\pi$ become a lower bound of the excitation continuum, while those in $q_x>0.7\pi$ keep an isolated mode.

In Fig. 4, the weights of the lowest excited states in $J_{cyc}/J_{\|}=0.1$ are compared with those in $J_{cyc}=0$. In $q_y=\pi$, the weights in $q_x \sim \pi$ are enhanced by $J_{cyc}$, while those in $q_x < \pi$ are suppressed. In $q_y=0$, on the contrary, the weights in $q_x > \pi/2$ are suppressed, while those in $q_x < \pi/2$ are enhanced. It is apparent that the distribution of the weights of DSF is drastically rearranged by a small cyclic four-spin interaction.

We next investigate dynamical properties of the systems described by the parameter set (ii) for ${\rm SrCu_2O_3}$ mentioned previously \cite{mizuno1}. 
We add the contribution of a diagonal interaction described as $J_{diag}\sum (\mbox{\boldmath$S$}_{1,i}\cdot\mbox{\boldmath$S$}_{2,i+1}+\mbox{\boldmath$S$}_{1,i+1}\cdot\mbox{\boldmath$S$}_{2,i})$ to the expressions (1) and (3). 
The results for $N=28$ are shown in Fig. 5. 
Note that the results for the parameter set (i) ($J_{\bot}/J_{\|}=0.5$) \cite{Johnson} have been already presented in Figs. 1 and 2. 
In $q_y=\pi$ and $0$, noticeable size dependence appears both in the poles and the residues. Therefore, the lowest excited states in $0 \leq q_x \leq \pi$ are the lower bounds of the excitation continuum as in the case of the parameter set (i).

In Fig. 6, the dispersion relations and the weights of the lowest excited states for both parameter sets (i) and (ii) are compared. In $q_y=\pi$, the dispersion relation for the parameter sets (ii) becomes flatter in $q_x<0.7\pi$, whereas the weights for both parameter sets are almost the same.  The results are qualitatively the same as those for $N=24$ \cite{mizuno2}. 
In $q_y=0$, the dispersion relation for the parameter set (ii) becomes flatter with suppressed weights in $q_x>\pi/2$. 
The difference in the distribution of the weights between two parameter sets may be difficult to observe experimentally. On the other hand, it has been shown that the effects of a flat dispersion relation of the lowest excited states appear in phonon-assisted optical absorption spectrums \cite{jb,legex,SCES}. To decide the proper model for ${\rm SrCu_2O_3}$, such experiments may be effective.

In summary, we have investigated dynamical properties of $S=1/2$ two-leg spin ladder systems. It has been shown that the largest weights in given $q_x$ lie in the lowest excited states. The distribution of the weights is influenced by a cyclic four-spin interaction. 
We have further shown that due to a cyclic four-spin interaction, the lowest excited states for $q_y=\pi$ in ${\rm La_{6}Ca_{8}Cu_{24}O_{41}}$ become a lower bound of the excitation continuum in $q_x<0.7\pi$, while those in $q_x>0.7\pi$ may form an isolated mode. 
We have also discussed the difference in dynamical properties between two systems proposed to describe ${\rm SrCu_2O_3}$. 
%
%
%


N.H. would like to thank K. Okunishi for useful comments. Our computational programs are based on TITPACK Ver. 2 by H. Nishimori. 
Numerical computation was partly carried out at the Yukawa Institute Computer Facility, Kyoto University, and the Supercomputer Center, the Institute for Solid State Physics, University of Tokyo. 
This work was partly supported by a Grant-in-Aid for Scientific Research from the Ministry of Education, Culture, Sports, Science, and Technology, Japan. 


%
%
%
%
%
%
%
%

\end{document}